\documentclass[aip,apl,twocolumn,noshowpacs,superscriptaddress,longbibliography]{revtex4}
\usepackage{graphicx}
\usepackage{amsmath,amssymb,amsfonts}
\usepackage[colorlinks,linkcolor=blue,urlcolor=blue,citecolor=blue]{hyperref}
\begin{document}
	
	
	\title{Measuring quantum geometric tensor of non-Abelian system in superconducting circuits}
	
	\author{Wen Zheng}
	\thanks{These authors contributed equally to this work.}
	\affiliation{National Laboratory of Solid State Microstructures, School of Physics, Nanjing University, Nanjing 210093, China}
	\author{Jianwen Xu}
	\thanks{These authors contributed equally to this work.}
	\affiliation{National Laboratory of Solid State Microstructures, School of Physics, Nanjing University, Nanjing 210093, China}
	
	\author{Zhuang Ma}
	\affiliation{National Laboratory of Solid State Microstructures, School of Physics, Nanjing University, Nanjing 210093, China}
	\author{Yong Li}
	\affiliation{National Laboratory of Solid State Microstructures, School of Physics, Nanjing University, Nanjing 210093, China}
	\author{Yuqian Dong}
	\affiliation{National Laboratory of Solid State Microstructures, School of Physics, Nanjing University, Nanjing 210093, China}
	\author{Yu Zhang}
	\affiliation{National Laboratory of Solid State Microstructures, School of Physics, Nanjing University, Nanjing 210093, China}
	\author{Xiaohan Wang}
	\affiliation{National Laboratory of Solid State Microstructures, School of Physics, Nanjing University, Nanjing 210093, China}

	\author{Guozhu Sun}
	\affiliation{School of Electronic Science and Engineering, Nanjing University, Nanjing 210093, China}
	\author{Peiheng Wu}
	\affiliation{School of Electronic Science and Engineering, Nanjing University, Nanjing 210093, China}	
	
	\author{Jie Zhao}
	\affiliation{National Laboratory of Solid State Microstructures, School of Physics, Nanjing University, Nanjing 210093, China}
	\author{Shaoxiong Li}
	\affiliation{National Laboratory of Solid State Microstructures, School of Physics, Nanjing University, Nanjing 210093, China}
	
	\author{Dong Lan}
	\email{land@nju.edu.cn}
	\affiliation{National Laboratory of Solid State Microstructures, School of Physics, Nanjing University, Nanjing 210093, China}
	\author{Xinsheng Tan}
	\email{tanxs@nju.edu.cn}
	\affiliation{National Laboratory of Solid State Microstructures, School of Physics, Nanjing University, Nanjing 210093, China}
	\author{Yang Yu}
	\email{yuyang@nju.edu.cn}
	\affiliation{National Laboratory of Solid State Microstructures, School of Physics, Nanjing University, Nanjing 210093, China}
	
	\date{\today}

	\begin{abstract}    
		{
			Topology played an important role in physics research during the last few decades. 
			In particular, the quantum geometric tensor that provides local information about topological properties has attracted much attention.
			It will reveal interesting topological properties in non-Abelian systems, which have not been realized in practice. Here, we use a four-qubit quantum system in superconducting circuits to construct a degenerate Hamiltonian with parametric modulation. By manipulating the Hamiltonian with periodic drivings, we simulate the Bernevig-Hughes-Zhang model and obtain the  quantum geometric tensor from interference oscillation.
			In addition, we reveal its topological feature by extracting the topological invariant, demonstrating an effective protocol for quantum simulation of non-Abelian system.                                                                                                                                                                                                                                                                                                                                                                                          
			
		}
	\end{abstract}
	

	\maketitle
	
	\textit{Introduction.}---
	Topology played a very important role in physics research in the last few decades \cite{nakahara2018geometry}, including the areas of high energy physics and condensed matter physics \cite{PhysRevD.12.3845, WU1976365, RevModPhys.82.3045, RevModPhys.83.1057, goldman_light_2014, RevModPhys.90.015001, zhang_topological_2018, RevModPhys.91.015006, buluta_quantum_2009, bliokh_quantum_2015}. 
	One of the most important quantities of topology is the quantum geometric tensor (QGT)  \cite{provost_riemannian_1980, PhysRevLett.99.095701, PhysRevB.81.245129}, of which an imaginary part is the Berry curvature, which is closely related to many physical phenomena such as the Aharonov-Bohm phase \cite{PhysRev.115.485} and topological orders in condensed matter \cite{PhysRevB.44.2664, RevModPhys.89.041004}.
	The QGT can be used to classify various topological materials. 
	Furthermore, the geometric phase is applied in the study of quantum computation and information \cite{ZANARDI199994, PhysRevA.61.010305, Bohm_gometric_phase_2003}, and it holds the promise to improve the performance of quantum gates due to its immunity to local fluctuations \cite{PhysRevLett.91.187902, PhysRevA.72.020301, PhysRevLett.102.030404, PhysRevA.96.052316}. 
	The real part of the QGT, called the quantum metric tensor, characterizes the distance between states in a Hilbert space, which is closely related to quantum fluctuations \cite{PhysRevD.23.357, PhysRevLett.72.3439}. 
	In the recent period of time, it has attracted interest of researchers in quantum phase transitions \cite{PhysRevE.74.031123, PhysRevLett.99.100603, sachdev_2011, carollo_geometry_2020} and quantum information theory \cite{PhysRevE.86.031137}.
	Both theoretical and experimental research has been conducted. 
	For instance, similar to the well-known Chern number that characterizes the topological properties of the Dirac monopole \cite{noauthor_quantised_nodate, RevModPhys.82.1959, ray_observation_2014}, the topological invariants of the tensor monopole are related to the geometric metric tensor \cite{PhysRevLett.121.170401}. 
	The four-dimensional matter predicted by theory has been simulated in superconducting circuits \cite{PhysRevLett.126.017702} and NV-center \cite{chen2021synthetic}, while its quantum phase transition is well characterized by the metric tensor \cite{PhysRevLett.121.170401}.   
	
	Compared to that of the Abelian quantum system, the quantum geometric tensor in the non-Abelian system has more complex mathematical structures and embodies more physics phenomena, including the four-dimensional Hall effect \cite{zhang2001four}.
	The topological order of the non-Abelian quantum systems can be extracted from the tensor, revealing the relationship between the physics phenomena and topological properties. 
	For instance, the second Chern number, as the topological invariant of the Yang monopole, is theoretically and experimentally studied \cite{yang_generalization_1978, sugawa_second_2018, PhysRevLett.117.015301, weisbrich_second_2021}.
	Furthermore, the Wilczek-Zee connection, which leads to the Holonomic properties in the generated system, allows the geometric gate to be widely used in quantum circuits \cite{PhysRevLett.52.2111, Duan_Geometric_2001, sugawa_wilson_2021, RevModPhys.80.1083}. 
	This subject is extensively and thoroughly studied for the optimization of quantum gates \cite{sjoqvist_non_adiabatic_2012, PhysRevLett.109.170501, PhysRevLett.102.070502, PhysRevLett.95.130501}. 
	
	Therefore, measuring the generalized non-Abelian quantum geometric tensor has both theoretical and practical significance.
	However, detecting the geometric tensor of a complex system is tricky in practice. 
	For an Abelian system, some approaches have been proposed, including modulation of Hamiltonian with sudden quench and periodic drivings \cite{PhysRevB.97.201117, PhysRevLett.122.210401, 10.1093/nsr/nwz193}. 
	These routines have been realized in various systems such as superconducting circuits \cite{PhysRevLett.122.210401} and NV-center \cite{10.1093/nsr/nwz193}. 
	For a non-Abelian system, several theoretical detection schemes have been proposed \cite{PhysRevResearch.3.033122}, but there is still a lack of experimental demonstration.          
	
	In this work, we construct a non-Abelian system in four-qubit superconducting circuits. Using longitudinal parametric modulation \cite{PhysRevApplied.6.064007, Matthew_2018, PhysRevApplied.13.064012},
	we demonstrate that the quantum geometric tensor, which depends on the geometry defined by a set of parameters, can be extracted by using geometric Rabi oscillations that are realized by periodically driving the system under a single parameter or under two parameters \cite{PhysRevB.97.201117, PhysRevResearch.3.033122}. We implement this protocol to simulate the Bernevig-Hughes-Zhang model and reveal its physics feature such as Z2 symmetry from the extracted topological invariant. The good agreement between experimental data and theoretical prediction confirms the validity of our approach.

	\begin{figure}
		\begin{minipage}[b]{0.5\textwidth}
			\centering
			\includegraphics[width=8.5cm]{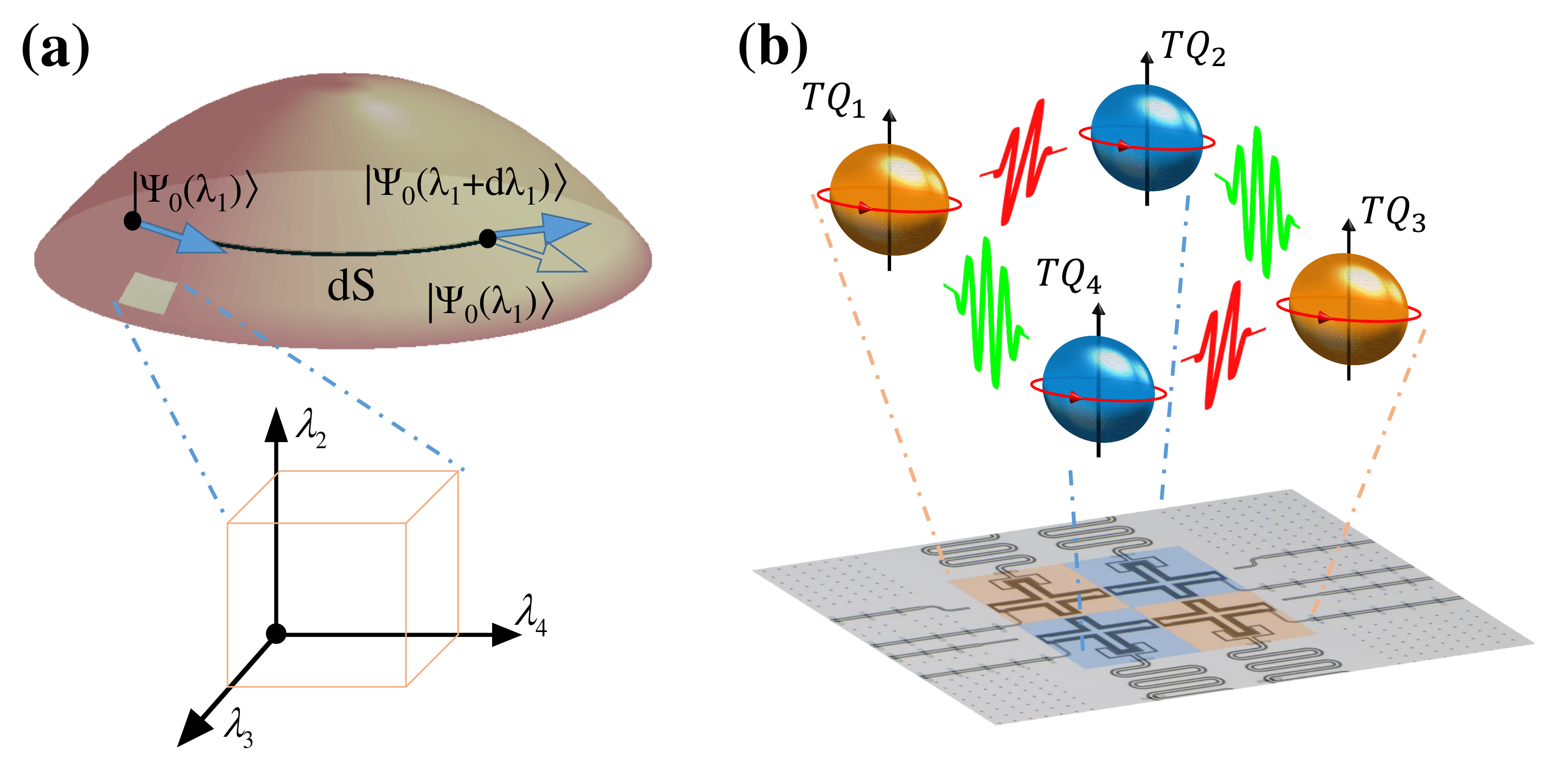}
		\end{minipage}
		\caption{
			\textbf{Illustration of the quantum geometric tensor in a non-Abelian system.}
			\textbf{(a)}$\,$ The QGT can be obtained from the distance between two nearby states in the four dimensional manifold spanned by the parameters $\lambda = (\lambda_1, \lambda_2, \lambda_3, \lambda_4)$.
			\textbf{(b)}$\,$ Schematic of our samples, which are four transmons capacitively coupled to each other.
			The simulated Hamiltonian can be realized by modulating the parameters of longitudinal driving applied to the superconducting circuits.
			\label{fig:FIG1}
		}
	\end{figure}
	\begin{figure}
		\begin{minipage}[b]{0.5\textwidth}
			\centering
			\includegraphics[width=8.5cm]{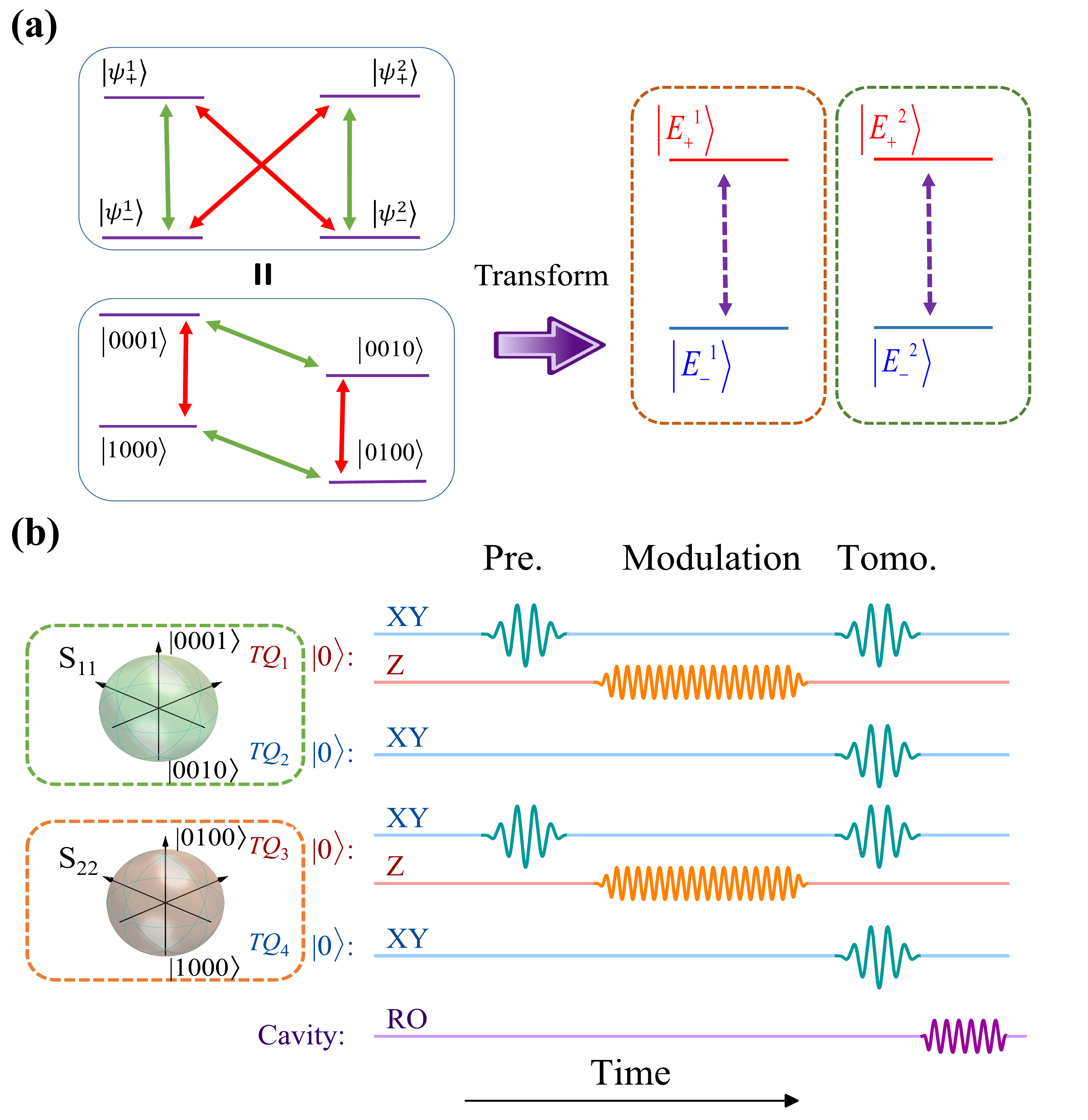}
		\end{minipage}
		\caption{
			\textbf{Non-Abelian system realized by longitudinal parametric modulation.}
			\textbf{(a)}$\,$ Flow chart of the Hamiltonian simulation.
			The original theoretical model can be equivalently mapped to the Hamiltonian constructed by the four-qubit system with longitudinal parametric modulation, which can be further simplified to the two independent subspaces under a basis transformation.
			\textbf{(b)}$\,$Pulse sequence.
			To construct the two-band two-degeneracy Hamiltonian as denoted in the dashed boxes, XY and Z designed pulses are applied.
			The protocol is realized by applying the designed flux pulses through the Z lines of tunable qubits $TQ_m$, which is followed by the quantum state tomography.
			\label{fig:Fig_nonAbelianSys}
		}
	\end{figure}
	\begin{figure*}
		\begin{minipage}[b]{1.0\textwidth}
			\centering
			\includegraphics[width=17cm]{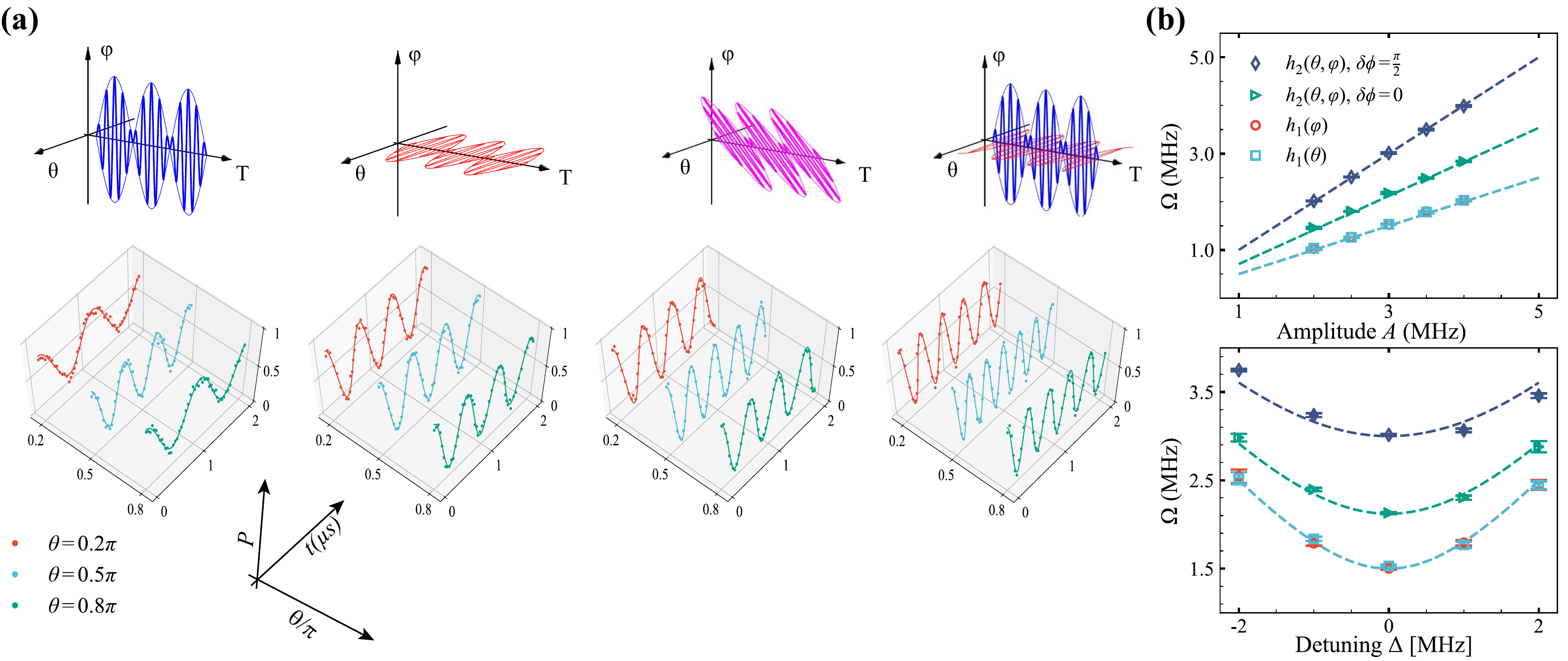}
		\end{minipage}
		\caption{
			\textbf{Rabi oscillation with different parametric modulation.}
			\textbf{(a)}$\,$ Top panel: Parametric modulation pulse applied to the subspace.
			From left to right: $\varphi$ modulation; $\theta$ modulation; $\theta$ and $\varphi$ modulation with $\delta\phi=0$; $\theta$ and $\varphi$ modulation with $\delta\phi=0.5\pi$.  
			Bottom panel: Measured Rabi oscillation at various ${\theta}$ obtained with corresponding routines of parametric modulation:  ${\theta}=0.2 \pi$, $0.5 \pi$, and $0.8 \pi$.
			The amplitude and frequency of the modulated pulse are $A/2\pi=3 \, \rm{MHz}$ and $\omega/2\pi=13 \, \rm{MHz}$, respectively. \textbf{(b)}$\,$  The Rabi oscillation frequency $\Omega$  in subspace $\mathcal{S}_{11}$ as a function of the driving amplitude $A$ and detuning $\Delta$, with $\theta=0.5\pi$ and $\omega = 13$ MHz. 
			The slope of each dashed line indicates the corresponding QGT according to Eq. \eqref{eq:QGTRabiOscillation}. Note that $h_1 (\theta)$ is almost overlaps with $h_1 (\varphi)$.
			\label{fig:Fig_RabiOsci}
		}
	\end{figure*}
	
	\textit{QGT in Non-Abelian systems.}---
	We consider an $N$-band Hamiltonian with $n$-degeneracy dimensions,
	which can be written as 
	$H_0(\lambda) = \sum_{\sigma=0}^{N} E_{\sigma} \sum_{i=1}^{n} |\psi_{i}^{\sigma}(\lambda) \rangle \langle \psi_{i}^{\sigma}(\lambda) |$ 
	with a set of dimensionless parameters
	$\lambda = (\lambda_1, \lambda_2, \dots)$, while $|\psi_{i}^{0}(\lambda) \rangle$ is the ground state. 
	Without loss of generality, we choose $N=1$ and $n=2$, so the dimension of the simulated Hamiltonian is four.
	Similar to that of the Abelian system, the geometric tensor can be derived from the distance between two nearby states in a Hilbert space, which is defined as $dS := \left\| |\Psi_0(\lambda_1 + d \lambda_1)\rangle -  |\Psi_0(\lambda_1)\rangle\right\| $, as illustrated in Fig. \ref{fig:FIG1}. 
	Here 
	$|\Psi_0(\lambda)\rangle = \sum_{i=1}^{n}|\psi_{i}^0 (\lambda)\rangle$.   
	The matrix element of the  QGT \cite{PhysRevB.81.245129}
	can be found as 
	\begin{equation}\label{eq:QGTDefinition}
	Q^{\mu \nu}_{i j} = \langle \partial_{\lambda_{\mu}} \psi_i^0 (\lambda) |[1 - \mathcal{P}(\lambda)]|\partial_{\lambda_{\nu}}\psi_j^0 (\lambda) \rangle,
	\end{equation}
	where 
	$\partial_{\lambda_{\mu, \nu}}\equiv\partial/\partial \lambda_{\mu,\nu}$ 
	and $\mathcal{P}(\lambda)$ stands for the projection operator, defined as $\mathcal{P}(\lambda) = \sum_{i=1}^{n}|\psi_{i}^0(\lambda)\rangle \langle \psi_{i}^0(\lambda) |$.
	Notice that the  QGT is a complex matrix, of which the real and imaginary parts are the Riemannian metric and Berry curvature respectively:
	$g^{\mu\nu} = (Q^{\mu\nu} + [Q^{\mu\nu}]^{\dagger})/2$, and 
	$F^{\mu\nu} = i(Q^{\mu\nu} - [Q^{\mu\nu}]^{\dagger})$.
	
	To measure the QGT of the degenerate system, 
	we construct the Hamiltonian $H_0$ and detect the interference oscillation with periodic weak parametric modulation \cite{ PhysRevResearch.3.033122}.
	The values of the metric tensor and Berry curvature can be extracted from the measured Rabi frequencies.
	In this routine, the QGT elements are obtained with one- and two-parameter modulation, respectively.
	With the modulating amplitude $A$ and the frequency $\omega$ satisfy $A \ll \omega$  \cite{PhysRevB.97.201117, PhysRevResearch.3.033122}, the corresponding Hamiltonians expanded to first order are  $H_0(\lambda) + h_1(\lambda_{\mu})$ 
	or $H_0(\lambda) + h_2(\lambda_{\mu}, \lambda_{\nu})$.
	Here $h_1(\lambda_{\mu}) = \frac{2A}{\omega} \cos{(\omega t + \phi_{\mu})} \partial_{\lambda_{\mu}} H_0 (\lambda)$ and
	$h_2(\lambda_{\mu}, \lambda_{\nu}) = \frac{2A}{\omega} [\cos{(\omega t+\phi_{\mu})} \partial_{\lambda_{\mu}} H_0 (\lambda) + \cos{(\omega t + \phi_{\nu})} \partial_{\lambda_{\nu}} H_0 (\lambda)]$.

	To simplify the experimental scheme and improve measurement fidelity, we rotate the Hamiltonian with a unitary transformation, so the coupling terms become zero \cite{supplementary3}. 
	In the modified Hamiltonian, the values of $ Q^{\mu \nu}_{i j}$ can be extracted from the oscillation frequencies of the new eigenstates \cite{PhysRevResearch.3.033122}.
	The relation between the value $Q_{QGT}$ and the Rabi oscillation is
	\begin{equation}\label{eq:QGTRabiOscillation}
	\Omega = \sqrt{A^2 Q_{QGT} + \Delta^2},
	\end{equation}
    for one- and two-parameter modulation $Q_{QGT}$ equals  
	$Q_{jj}^{\nu \nu}$ and $Q_{jj}^{\mu \mu} + Q_{jj}^{\nu \nu} + e^{-i\delta\phi}Q_{jj}^{\mu \nu} + e^{i\delta\phi}Q_{jj}^{\nu \mu}$ respectively.
	Here, $\delta \phi = \phi_{\nu} - \phi_{\mu}$ is the phase difference between the two periodic modulation pulses, $\Delta$ is the detuning between the gap of eigenenergy levels and the driving frequency $\omega$.

	\textit{Build non-Abelian system.}---
	The Hamiltonian constructed by four qubits can be written as
	$H_s = \sum_{k} \omega_{k}a^{\dagger}_k a_k + \frac{\alpha_k}{2}a^{\dagger}_k a^{\dagger}_k a_k a_k + \sum_{kl} J_{kl}(a_k + a_k^{\dagger})(a_l + a_l^{\dagger})$
	with indices $k\in \{1,\,2,\,3,\,4\}$ and $kl \in \{12,\,23,\,34,\,41\}$ to mark qubits.
	Here, $a_k$ ($a^{\dagger}_k$) is the annihilation (creation) operator
	and $J_{kl}$ is the nearest-neighbor coupling strength, while $\alpha_k$ denotes the anharmonicity of the $k$-th transmon, of which the frequency in the lowest two levels is labeled as $\omega_{k}$ \cite{supplementary1}.
	To realize the desired degenerate Hamiltonian $H_0(\lambda)$, whose level structure can be equivalent to a diamond shape as shown in the right panel of Fig. \ref{fig:Fig_nonAbelianSys}(a),
	we simultaneously introduce the longitudinal parametric modulation pulses
	$\omega_{m}(t) = \bar{\omega}_m + \omega^{T}_{m1}\cos{(2 \pi f_{m1}  t+\beta_{m1})} +  \omega^{T}_{m2}\cos{(2 \pi f_{m2} t+{\beta}_{m2})}$ by applying flux pulses to the tunable qubit $QT_m$ \cite{supplementary2}.	
	$\bar{\omega}_m$ is the mean frequency of $TQ_m$ during the longitudinal parametric modulation,
	$\omega^{T}_{mi}$,  $f_{mi}$ and $\beta_{mi}$ is the amplitude, frequency and phase of the applied longitudinal pulse. The two-tune pulses of parametric modulation realize the diamond-shape couplings as illustrated in Fig. \ref{fig:FIG1}(b). In this approach the parameters of each coupling term can be freely adjusted, making that
	the general Non-Abelian Hamiltonian  as shown in Fig. \ref{fig:Fig_nonAbelianSys}(a) can be directly realized in the Hilbert space spanned by $\{|0001\rangle,\,|0010\rangle,\,|0100\rangle,\,|1000\rangle\}$ \cite{supplementary2}.
	
	\textit{Experimental results.}---
	In practice, we detect the specific metrics to study the topological properties of physical models, here we constructed the Bernevig-Hughes-Zhang (BHZ) model as an example to demonstrate this scheme \cite{bernevig_quantum_2006, PhysRevLett.127.136802, supplementary3}. The BHZ model is closely related to the spin-Hall effect, which can be characterized by the spin Chern numbers.  The Hamiltonian is 
	\begin{equation}\label{eq:BHZ_model}
	\begin{aligned}
	H_{BHZ} =
	\left[
	\begin{matrix}
	B_{z} & 0 & B_{x} - i B_{y} & B_{g}\\
	0 & B_{z} & B_{g} & - B_{x} - i B_{y}\\
	B_{x} + i B_{y} & B_{g} & - B_{z} & 0\\
	B_{g} & - B_{x} + i B_{y} & 0 & - B_{z}
	\end{matrix}
	\right],
	\end{aligned}
	\end{equation}
	where  $B_{x} = H_{xy} \sin{k_x}, B_{y} =  H_{xy} \sin{k_y},B_{z} = M - 2 H_{z}(2 - \cos{k_x} - \cos{k_y})$.
	The momentum $k_x, \, k_y$ belong to $[-\pi, \pi]$, while the parameters $H_{xy}$, $H_{z}$, and $B_{g}$ depend on the quantum well geometry, $M$ is related to the phase transition.
	In our superconducting circuit we can realize this BHZ Hamiltonian, which is a non-Abelian form in a four-dimensional parameter space. 
	Its entire components of the QGT can be obtained by our approach \cite{supplementary4}, while it is not necessary in practice. The BHZ model is equivalent to two copies of the Haldane Model \cite{PhysRevLett.61.2015, PhysRevLett.127.136802}, which can be mapped to the Hilbert space of two degenerate systems.
	Therefore,  the original Hamiltonian  can be deformed to the individual degenerate subspace by a unitary transformation as shown in Fig. \ref{fig:Fig_nonAbelianSys}(a).
	The parametric modulation in experiment can be simplified as $\omega_{1}(t) = \bar{\omega}_1 + \omega^{T}_{1} \cos{ (2 \pi f_{1} t+\varphi)}$ and $\omega_{3}(t) =  \bar{\omega}_3 +  \omega^{T}_{3} \cos{(2 \pi f_{3} t-\varphi)}$, leading that
	the system Hamiltonian in spherical coordinate can be written as \cite{supplementary2}
	\begin{equation}\label{eq:H_na}
	\begin{aligned}
	{H}
	&= {\Omega_0}(\cos{\theta}|0001\rangle \langle 0001| - \cos{\theta}|0010\rangle \langle 0010| \\
	&+ \cos{\theta}|0100\rangle \langle 0100| - \cos{\theta}|1000\rangle \langle 1000| \\
	&+ \sin{\theta}e^{i\varphi}|0010\rangle \langle 0001| - \sin{\theta}e^{-i\varphi}|1000\rangle \langle 0100| + h.c.),\\
	\end{aligned}
	\end{equation}
	where $\Omega_0=\sqrt{\Omega^2 + \Delta^2}$. 
	Here $\Omega$ is the Rabi oscillation frequency in Eq. (\ref{eq:QGTRabiOscillation}), with  $\Omega=|J_{12}   \mathcal{J}_{1}(\frac{\omega_{1}^{T}}{2 \pi f_{1}})|=|J_{34} \mathcal{J}_{1}(\frac{\omega_{3}^{T}}{2 \pi f_{3}})|$.  $\mathcal{J}_1$ denotes the first kind 1st-order Bessel function. 
	The spherical coordinate $\theta= \arctan({\Omega}/{\Delta})$. 
	The QGT measurement, which contains initial state preparation, parametric modulation, and quantum state tomography, is demonstrated on four qubits by modulating the controlling parameters $\theta$ and $\varphi$.
	
\begin{figure}
		\begin{minipage}[b]{0.5\textwidth}
			\centering
			\includegraphics[width=8cm]{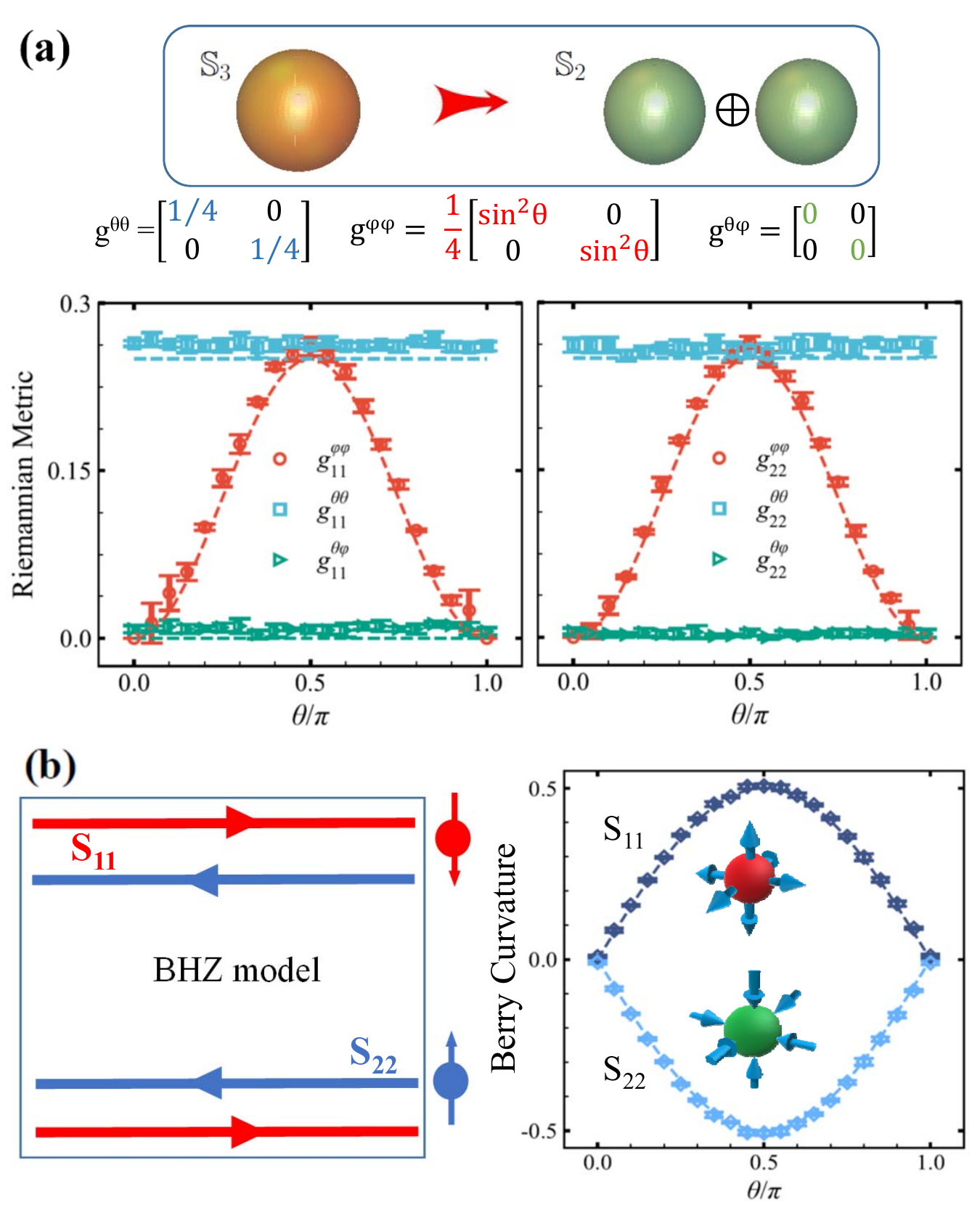}
		\end{minipage}
		\caption{
			\textbf{ Extraction of the quantum geometric tensor from Rabi oscillation.}
			\textbf{(a)}$\,$  Top panel: Under the MS transformation, the manifold of the BHZ Hamiltonian transform from a sphere in a four-dimensional parameter space to a direct sum of two $\mathbb{S}_2$ spheres.
			Therefore, each measurement obtained by our approach is part of the original quantum geometric metric tensor. Bottom panel: The measured Riemannian metric, data in different colors correspond to different matrix elements. 
			\textbf{(b)} $\,$ Left: The illustration of the BHZ model, the spin up (down) charge carriers correspond to the measured Berry curvature of subspace $\mathcal{S}_{11}$ ($\mathcal{S}_{22}$) shown in the right panel.
			The experimental data (solid points) agreed well with theoretical predictions (dashed lines).
			The error bars indicate standard deviation.
			\label{fig:Fig_NAQGT}
		}
	\end{figure}

	To obtain the complete quantum metric of the manifold, we need to execute the procedure by preparing different initial states throughout the entire Hilbert space. To improve measurement fidelity, we use a unitary transformation to rotate the frame axis.
	The eigenstates of the initial Hamiltonian remain at $|0001\rangle$ and $|0100\rangle$ \cite{PhysRevLett.122.210401}, making the modulation pulse significantly simplified.
	With this routine, the weak periodic driving term $h_1(\lambda_{\mu})$ 
	and $h_2(\lambda_{\mu}, \lambda_{\nu})$ can be realized by accurately designing the frequency and amplitude of the flux pulses  \cite{supplementary2}. 
	For instance, in the measurement of $g^{\varphi\varphi}_{11}$, The parameter ${\varphi}$ in the modulated pulse  simplified as $2A/\omega\cos(\omega t)$  with $A/2\pi=3$ MHz and $\omega/2\pi = 13$ MHz.
	Fig. \ref{fig:Fig_RabiOsci} shows the pulse schemes and measured Rabi patterns, with ${\theta}=0.2 \pi$, $0.5 \pi$ and $0.8 \pi$, respectively.
	Similarly, the Rabi patterns of the other QGT terms ($g^{\theta\theta}_{11}$, $g^{\varphi\theta}_{11}$, and  $F^{\varphi\theta}_{11}$) can also be obtained.
	Furthermore, we can verify the relationship between the QGT and the parameters of the periodic weak field in Eq. (\ref{eq:QGTRabiOscillation}).
	As shown in Fig. \ref{fig:Fig_RabiOsci}(b), we vary the amplitude $A$ and detuning $\Delta$ as a double-check.
	The data demonstrated in the figure are measured at ${\theta} = 0.5\pi$.
	Based on Eq. \eqref{eq:QGTRabiOscillation}, in the top panel of Fig. \ref{fig:Fig_RabiOsci}(b), there is a linear relationship between $A$ and $\Omega$ if $\Delta/2\pi = 0\, \rm{MHz}$, so the values of the QGT can be extracted from the experimental results, in the bottom panel of Fig. \ref{fig:Fig_RabiOsci}(b), by setting $A/2\pi=3\, \rm{MHz}$, the relationship between $\Delta$ and $\Omega$ also agrees well with the predictions (dashed lines).
	Note that $h_1 (\theta)$ is almost identical to $h_1 (\varphi)$.
	With this modulating protocol, we can extract the QGT of the simulating Hamiltonian, revealing the corresponding geometric properties. As
	shown in Fig. \ref{fig:Fig_NAQGT}(a), The experimental $Q_{\mu\nu}$ are in good agreement with theoretical predictions, which are 
	\begin{equation}
	\begin{aligned}
	& Q^{\theta\theta} = \frac{1}{4}\left[\begin{aligned} &1 & 0\\ &0 & 1\\\end{aligned}\right],
	& Q^{\varphi\varphi} = \frac{1}{4}\sin^2 \theta \left[\begin{aligned} &1& 0\\ &0& 1\\\end{aligned}\right],\\
	& Q^{\theta\varphi} = \frac{i}{4}\sin \theta \left[\begin{aligned} &1& 0\\ &0& -1\\\end{aligned}\right]
	\end{aligned}
	\end{equation} 
	It is worth noting that the Morris-Shore (MS) transformation used in our scheme reduces the parameter dimension, hence the complete components of the original geometric tensor are obtained by repeating this approach with different unitary transformation \cite{supplementary4}. 
	
	Furthermore, we study the topological invariant in the BHZ model, which is the spin Chern number. The spin up and down charge carriers correspond to the individual subspace  $\mathcal{S}_{11}$ and $\mathcal{S}_{22}$ we constructed, of which the corresponding topological invariant is the first Chern number $\mathcal{C_{\pm}}$.  Since the manifold is covered by three dimensional sphere surface $\mathbb{S}_2$, in spherical coordinates the Chern number of individual subspace can be written as
	\begin{equation}\label{eq:Chern_F}
	\mathcal{C} = \frac{1}{2\pi} \int_{\mathbb{S}_2} F_{\theta\phi} d\theta d\phi.
	\end{equation}
	The relation  $\sqrt{\rm{det}\,g}=\frac{1}{2}|F_{\theta\phi}|$ \cite{PhysRevLett.121.170401, zhang2021revealing, PhysRevB.104.195133, mera_relating_2022} ensures that we can obtain the Chern number from the metric tensor $g^{\varphi\varphi}_{11(22)}$, $g^{\theta\theta}_{11(22)}$, $g^{\theta\varphi}_{11(22)}$ and Berry curvature $F^{\theta\varphi}_{11(22)}$ we measured, respectively. 
	The accumulated values of the subspace $\mathcal{S}_{11}$ and $\mathcal{S}_{22}$ extracted from the Berry curvature are $\mathcal{C_+} = 1.019 \pm 0.005$ and $\mathcal{C_-} = -1.003 \pm 0.010$, which correspond to the positive and negative topological charges respectively, as illustrated in Fig. \ref{fig:Fig_NAQGT}(b).  The spin Chern number is obtained as $\mathcal{C}_{scn}=(\mathcal{C}_+-\mathcal{C}_-)/2 = 1.011 \pm 0.005$ while $\mathcal{C}_++\mathcal{C}_- = 0.016 \pm 0.013$ due to the time-reversal symmetry,  indicating the Z2 symmetry of the Bernevig-Hughes-Zhang model \cite{PhysRevLett.95.146802, PhysRevLett.95.226801, PhysRevLett.95.136602, PhysRevLett.97.036808, sheng_spin_2013}.  Therefore, the topological properties of BHZ model  can be explored directly with our approach.
	
	\textit{Conclusions.}---
	In this article, we have demonstrated the simulation of a non-Abelian system in superconducting circuits.
	Using periodic driving, we measure the geometric tensor of this quantum system.  
	In addition, we extract the topological invariant from the measured metric tensor and the Berry curvature, respectively, proving the feasibility and validity of our routine. 
	Furthermore, our scheme is universal and can be extended to quantum simulations of other models.                                                                        
	
	~\\ 
	\textbf{Acknowledgement.}---
	This work was supported by
	the Key R\&D Program of Guangdong Province (Grant No. 2018B030326001),
	NSFC (Grants No. 11474152, No. 12074179, No. U21A20436, and No. 61521001), NSF of Jiangsu Province (Grant No. BE2021015-1).
	

\begin{thebibliography}{10}
		\providecommand{\url}[1]{\texttt{#1}}
		\providecommand{\urlprefix}{}
		\providecommand{\eprint}[2][]{\url{#2}}
		
		\bibitem{nakahara2018geometry}
		M.~Nakahara.
		\newblock \emph{Geometry, topology and physics}.
		\newblock CRC press (2018).
		
		\bibitem{PhysRevD.12.3845}
		T.~T. Wu and C.~N. Yang,
		  \href{http://dx.doi.org/10.1103/PhysRevD.12.3845}{Phys. Rev. D \textbf {12}
		  3845,  (1975)}.
		
		\bibitem{WU1976365}
		T.~T. Wu and C.~N. Yang,
		  \href{http://dx.doi.org/https://doi.org/10.1016/0550-3213(76)90143-7}{Nucl.
		  Phys. B \textbf {107} 365,  (1976)}.
		
		\bibitem{RevModPhys.82.3045}
		M.~Z. Hasan and C.~L. Kane,
		  \href{http://dx.doi.org/10.1103/RevModPhys.82.3045}{Rev. Mod. Phys. \textbf
		  {82} 3045,  (2010)}.
		
		\bibitem{RevModPhys.83.1057}
		X.-L. Qi and S.-C. Zhang,
		  \href{http://dx.doi.org/10.1103/RevModPhys.83.1057}{Rev. Mod. Phys. \textbf
		  {83} 1057,  (2011)}.
		
		\bibitem{goldman_light_2014}
		N.~Goldman, G.~Juzeliūnas, P.~Öhberg, and I.~B. Spielman,
		  \href{http://dx.doi.org/10.1088/0034-4885/77/12/126401}{Rep. Prog. Phys.
		  \textbf {77} 126401,  (2014)}.
		
		\bibitem{RevModPhys.90.015001}
		N.~P. Armitage, E.~J. Mele, and A.~Vishwanath,
		  \href{http://dx.doi.org/10.1103/RevModPhys.90.015001}{Rev. Mod. Phys. \textbf
		  {90} 015001,  (2018)}.
		
		\bibitem{zhang_topological_2018}
		D.-W. Zhang, Y.-Q. Zhu, Y.~X. Zhao, H.~Yan, and S.-L. Zhu,
		  \href{http://dx.doi.org/10.1080/00018732.2019.1594094}{Adv. Phys. \textbf
		  {67} 253,  (2018)}.
		
		\bibitem{RevModPhys.91.015006}
		T. Ozawa, H.~M. Price, A. Amo, N. Goldman, M. Hafezi, L. Lu, M.~C. Rechtsman, D. Schuster, J. Simon, O. Zilberberg, and I. Carusotto,
		  \href{http://dx.doi.org/10.1103/RevModPhys.91.015006}{Rev. Mod. Phys. \textbf
		  {91} 015006,  (2019)}.
		
		\bibitem{buluta_quantum_2009}
		I.~Buluta and F.~Nori, \href{http://dx.doi.org/10.1126/science.1177838}{Science
		  \textbf {326} 108,  (2009)}.
		
		\bibitem{bliokh_quantum_2015}
		K.~Y. Bliokh, D.~Smirnova, and F.~Nori,
		  \href{http://dx.doi.org/10.1126/science.aaa9519}{Science \textbf {348} 1448,
		  (2015)}.
		
		\bibitem{provost_riemannian_1980}
		J.~P. Provost and G.~Vallee,
		  \href{http://dx.doi.org/10.1007/BF02193559}{Commun. Math. Phys. \textbf {76}
		  289,  (1980)}.
		
		\bibitem{PhysRevLett.99.095701}
		L.~Campos~Venuti and P.~Zanardi,
		  \href{http://dx.doi.org/10.1103/PhysRevLett.99.095701}{Phys. Rev. Lett.
		  \textbf {99} 095701,  (2007)}.
		
		\bibitem{PhysRevB.81.245129}
		Y.-Q. Ma, S.~Chen, H.~Fan, and W.-M. Liu,
		  \href{http://dx.doi.org/10.1103/PhysRevB.81.245129}{Phys. Rev. B \textbf {81}
		  245129,  (2010)}.
		
		\bibitem{PhysRev.115.485}
		Y.~Aharonov and D.~Bohm, \href{http://dx.doi.org/10.1103/PhysRev.115.485}{Phys.
		  Rev. \textbf {115} 485,  (1959)}.
		
		\bibitem{PhysRevB.44.2664}
		X.~G. Wen, \href{http://dx.doi.org/10.1103/PhysRevB.44.2664}{Phys. Rev. B
		  \textbf {44} 2664,  (1991)}.
		
		\bibitem{RevModPhys.89.041004}
		X.-G. Wen, \href{http://dx.doi.org/10.1103/RevModPhys.89.041004}{Rev. Mod.
		  Phys. \textbf {89} 041004,  (2017)}.
		
		\bibitem{ZANARDI199994}
		P.~Zanardi and M.~Rasetti,
		  \href{http://dx.doi.org/https://doi.org/10.1016/S0375-9601(99)00803-8}{Phys.
		  Lett. A \textbf {264} 94,  (1999)}.
		
		\bibitem{PhysRevA.61.010305}
		J.~Pachos, P.~Zanardi, and M.~Rasetti,
		  \href{http://dx.doi.org/10.1103/PhysRevA.61.010305}{Phys. Rev. A \textbf {61}
		  010305(R),  (1999)}.
		
		\bibitem{Bohm_gometric_phase_2003}
		A.~Bohm, A.~Mostafazadeh, H.~Koizumi, Q.~Niu, and J.~Zwanziger.
		\newblock \emph{The Geometric Phase in Quantum Systems: Foundations,
		  Mathematical Concepts, and Applications in Molecular and Condensed Matter
		  Physics} (2003).
		
		\bibitem{PhysRevLett.91.187902}
		S.-L. Zhu and Z.~D. Wang,
		  \href{http://dx.doi.org/10.1103/PhysRevLett.91.187902}{Phys. Rev. Lett.
		  \textbf {91} 187902,  (2003)}.
		
		\bibitem{PhysRevA.72.020301}
		S.-L. Zhu and P.~Zanardi,
		  \href{http://dx.doi.org/10.1103/PhysRevA.72.020301}{Phys. Rev. A \textbf {72}
		  020301(R),  (2005)}.
		
		\bibitem{PhysRevLett.102.030404}
		S.~Filipp, J.~Klepp, Y.~Hasegawa, C.~Plonka-Spehr, U.~Schmidt, P.~Geltenbort, and H.~Rauch,
		  \href{http://dx.doi.org/10.1103/PhysRevLett.102.030404}{Phys. Rev. Lett.
		  \textbf {102} 030404,  (2009)}.
		
		\bibitem{PhysRevA.96.052316}
		P.~Z. Zhao, X.-D. Cui, G.~F. Xu, E.~Sj\"oqvist, and D.~M. Tong,
		  \href{http://dx.doi.org/10.1103/PhysRevA.96.052316}{Phys. Rev. A \textbf {96}
		  052316,  (2017)}.
		
		\bibitem{PhysRevD.23.357}
		W.~K. Wootters, \href{http://dx.doi.org/10.1103/PhysRevD.23.357}{Phys. Rev. D
		  \textbf {23} 357,  (1981)}.
		
		\bibitem{PhysRevLett.72.3439}
		S.~L. Braunstein and C.~M. Caves,
		  \href{http://dx.doi.org/10.1103/PhysRevLett.72.3439}{Phys. Rev. Lett. \textbf
		  {72} 3439,  (1994)}.
		
		\bibitem{PhysRevE.74.031123}
		P.~Zanardi and N.~Paunkovi\ifmmode~\acute{c}\else \'{c}\fi{},
		  \href{http://dx.doi.org/10.1103/PhysRevE.74.031123}{Phys. Rev. E \textbf {74}
		  031123,  (2006)}.
		
		\bibitem{PhysRevLett.99.100603}
		P.~Zanardi, P.~Giorda, and M.~Cozzini,
		  \href{http://dx.doi.org/10.1103/PhysRevLett.99.100603}{Phys. Rev. Lett.
		  \textbf {99} 100603,  (2007)}.
		
		\bibitem{sachdev_2011}
		S.~Sachdev.
		\newblock \emph{Quantum Phase Transitions}.
		\newblock Cambridge University Press, 2 edition (2011).
		
		\bibitem{carollo_geometry_2020}
		A.~Carollo, D.~Valenti, and B.~Spagnolo,
		  \href{http://dx.doi.org/10.1016/j.physrep.2019.11.002}{Phys. Rep. \textbf
		  {838} 1,  (2020)}.
		
		\bibitem{PhysRevE.86.031137}
		A.~Dey, S.~Mahapatra, P.~Roy, and T.~Sarkar,
		  \href{http://dx.doi.org/10.1103/PhysRevE.86.031137}{Phys. Rev. E \textbf {86}
		  031137,  (2012)}.
		
		\bibitem{noauthor_quantised_nodate}
		P.~A.~M. Dirac, \href{https://doi.org/10.1098/rspa.1931.0130}{Proc. R. Soc.
		  Lond. A \textbf {133} 60}.
		
		\bibitem{RevModPhys.82.1959}
		D.~Xiao, M.-C. Chang, and Q.~Niu,
		  \href{http://dx.doi.org/10.1103/RevModPhys.82.1959}{Rev. Mod. Phys. \textbf
		  {82} 1959,  (2010)}.
		
		\bibitem{ray_observation_2014}
		M.~W. Ray, E.~Ruokokoski, S.~Kandel, M.~Möttönen, and D.~S. Hall,
		  \href{http://dx.doi.org/10.1038/nature12954}{Nature \textbf {505} 657,
		  (2014)}.
		
		\bibitem{PhysRevLett.121.170401}
		G.~Palumbo and N.~Goldman,
		  \href{http://dx.doi.org/10.1103/PhysRevLett.121.170401}{Phys. Rev. Lett.
		  \textbf {121} 170401,  (2018)}.
		
		\bibitem{PhysRevLett.126.017702}
		X. Tan, D.-W. Zhang, W. Zheng, X. Yang, S. Song, Z. Han, Y. Dong, Z. Wang, D. Lan, H. Yan, S.-L. Zhu, and Y. Yu,
		  \href{http://dx.doi.org/10.1103/PhysRevLett.126.017702}{Phys. Rev. Lett.
		  \textbf {126} 017702,  (2021)}.
		
		\bibitem{chen2021synthetic}
		M.~Chen, C.~Li, G.~Palumbo, Y.-Q. Zhu, N.~Goldman, and P.~Cappellaro,
		\href{http://dx.doi.org/10.1126/science.abe6437}{Science \textbf {375} 1017,
		(2022)}.
		
		\bibitem{zhang2001four}
		S.-C. Zhang and J.~Hu, {Science \textbf {294} 823,  (2001)}.
		
		\bibitem{yang_generalization_1978}
		C.~N. Yang, {J. Math. Phys. \textbf {19} 10,  (1978)}.
		
		\bibitem{sugawa_second_2018}
		S.~Sugawa, F.~Salces-Carcoba, A.~R. Perry, Y.~Yue, and I.~B. Spielman,
		  \href{http://dx.doi.org/10.1126/science.aam9031}{Science \textbf {360} 1429,
		  (2018)}.
		
		\bibitem{PhysRevLett.117.015301}
		M.~Kolodrubetz, \href{http://dx.doi.org/10.1103/PhysRevLett.117.015301}{Phys.
		  Rev. Lett. \textbf {117} 015301,  (2016)}.
		
		\bibitem{weisbrich_second_2021}
		H.~Weisbrich, R.~Klees, G.~Rastelli, and W.~Belzig,
		  \href{http://dx.doi.org/10.1103/PRXQuantum.2.010310}{PRX Quantum \textbf {2}
		  010310,  (2021)}.
		
		\bibitem{PhysRevLett.52.2111}
		F.~Wilczek and A.~Zee,
		  \href{http://dx.doi.org/10.1103/PhysRevLett.52.2111}{Phys. Rev. Lett. \textbf
		  {52} 2111,  (1984)}.
		
		\bibitem{Duan_Geometric_2001}
		L.-M. Duan, J.~I. Cirac, and P.~Zoller,
		  \href{http://dx.doi.org/10.1126/science.1058835}{Science \textbf {292} 1695,
		  (2001)}.
		
		\bibitem{sugawa_wilson_2021}
		S.~Sugawa, F.~Salces-Carcoba, Y.~Yue, A.~Putra, and I.~B. Spielman,
		  \href{http://dx.doi.org/10.1038/s41534-021-00483-2}{npj Quantum Inform.
		  \textbf {7} 144,  (2021)}.
		
		\bibitem{RevModPhys.80.1083}
		C.~Nayak, S.~H. Simon, A.~Stern, M.~Freedman, and S.~Das~Sarma,
		  \href{http://dx.doi.org/10.1103/RevModPhys.80.1083}{Rev. Mod. Phys. \textbf
		  {80} 1083,  (2008)}.
		
		\bibitem{sjoqvist_non_adiabatic_2012}
		E.~Sjöqvist, D.~M. Tong, L.~Mauritz~Andersson, B.~Hessmo, M.~Johansson, and
		  K.~Singh, \href{http://dx.doi.org/10.1088/1367-2630/14/10/103035}{New J.
		  Phys. \textbf {14} 103035,  (2012)}.
		
		\bibitem{PhysRevLett.109.170501}
		G.~F. Xu, J.~Zhang, D.~M. Tong, E.~Sj\"oqvist, and L.~C. Kwek,
		  \href{http://dx.doi.org/10.1103/PhysRevLett.109.170501}{Phys. Rev. Lett.
		  \textbf {109} 170501,  (2012)}.
		
		\bibitem{PhysRevLett.102.070502}
		O.~Oreshkov, T.~A. Brun, and D.~A. Lidar,
		  \href{http://dx.doi.org/10.1103/PhysRevLett.102.070502}{Phys. Rev. Lett.
		  \textbf {102} 070502,  (2009)}.
		
		\bibitem{PhysRevLett.95.130501}
		L.-A. Wu, P.~Zanardi, and D.~A. Lidar,
		  \href{http://dx.doi.org/10.1103/PhysRevLett.95.130501}{Phys. Rev. Lett.
		  \textbf {95} 130501,  (2005)}.
		
		\bibitem{PhysRevB.97.201117}
		T.~Ozawa and N.~Goldman,
		  \href{http://dx.doi.org/10.1103/PhysRevB.97.201117}{Phys. Rev. B \textbf {97}
		  201117(R),  (2018)}.
		
		\bibitem{PhysRevLett.122.210401}
		X. Tan, D.~W. Zhang, Z. Yang, J. Chu, Y.~Q. Zhu, D. Li, X. Yang, S. Song, Z. Han, Z. Li, Y. Dong, H.~F. Yu, H. Yan, S.~L. Zhu, and Y. Yu,
		  \href{http://dx.doi.org/10.1103/PhysRevLett.122.210401}{Phys. Rev. Lett.
		  \textbf {122} 210401,  (2019)}.
		
		\bibitem{10.1093/nsr/nwz193}
		M.~Yu \emph{et~al.}, \href{http://dx.doi.org/10.1093/nsr/nwz193}{Natl. Sci.
		  Rev. \textbf {7} 254,  (2019)}.
		
		\bibitem{PhysRevResearch.3.033122}
		H.~Weisbrich, G.~Rastelli, and W.~Belzig,
		  \href{http://dx.doi.org/10.1103/PhysRevResearch.3.033122}{Phys. Rev. Research
		  \textbf {3} 033122,  (2021)}.
		
		\bibitem{PhysRevApplied.6.064007}
		D.~C. McKay, S.~Filipp, A.~Mezzacapo, E.~Magesan, J.~M. Chow, and J.~M.
		  Gambetta, \href{http://dx.doi.org/10.1103/PhysRevApplied.6.064007}{Phys. Rev.
		  Applied \textbf {6} 064007,  (2016)}.
		
		\bibitem{Matthew_2018}
		M.~Reagor \emph{et~al.}, \href{http://dx.doi.org/10.1126/sciadv.aao3603}{Sci.
		  Adv. \textbf {4} eaao3603,  (2018)}.
		
		\bibitem{PhysRevApplied.13.064012}
		J. Chu, D. Li, X. Yang, S. Song, Z. Han, Z. Yang, Y. Dong, W. Zheng, Z. Wang, X. Yu, D. Lan, X. Tan, and Y. Yu,
		  \href{http://dx.doi.org/10.1103/PhysRevApplied.13.064012}{Phys. Rev. Applied
		  \textbf {13} 064012,  (2020)}.
		
		\bibitem{supplementary3}
		Please see the Supplementary Material for the diagonalization and measurement
		  of non-Abelian quantum geometric tensor.
		
		\bibitem{supplementary1}
		Please see the Supplementary Material for the sample information and
		  experimental setup.
		
		\bibitem{supplementary2}
		Please see the Supplementary Material for the realization of the Non-Abelian
		  system in superconducting circuits.
		
		\bibitem{bernevig_quantum_2006}
		B.~A. Bernevig, T.~L. Hughes, and S.-C. Zhang,
		  \href{http://dx.doi.org/10.1126/science.1133734}{Science \textbf {314} 1757,
		  (2006)}.
		
		\bibitem{PhysRevLett.127.136802}
		Q.-X. Lv, Y.-X. Du, Z.-T. Liang, H.-Z. Liu, J.-H. Liang, L.-Q. Chen, L.-M. Zhou, S.-C. Zhang, D.-W. Zhang, B.-Q. Ai, H. Yan, and S.-L. Zhu,
		  \href{http://dx.doi.org/10.1103/PhysRevLett.127.136802}{Phys. Rev. Lett.
		  \textbf {127} 136802,  (2021)}.
		
		\bibitem{supplementary4}
		Please see the Supplementary Material for the BHZ model.
		
		\bibitem{PhysRevLett.61.2015}
		F.~D.~M. Haldane, \href{http://dx.doi.org/10.1103/PhysRevLett.61.2015}{Phys.
		  Rev. Lett. \textbf {61} 2015,  (1988)}.
		
		\bibitem{zhang2021revealing}
		A.~Zhang, \href{http://dx.doi.org/10.1088/1674-1056/ac2f2c}{Chin. Phys. B
		  \textbf {31} 040201,  (2022)}.
		
		\bibitem{PhysRevB.104.195133}
		G.~von Gersdorff and W.~Chen,
		  \href{http://dx.doi.org/10.1103/PhysRevB.104.195133}{Phys. Rev. B \textbf
		  {104} 195133,  (2021)}.
		
		\bibitem{mera_relating_2022}
		B.~Mera, A.~Zhang, and N.~Goldman,
		  \href{http://dx.doi.org/10.21468/SciPostPhys.12.1.018}{SciPost Phys. \textbf
		  {12} 018,  (2022)}.
		
		\bibitem{PhysRevLett.95.146802}
		C.~L. Kane and E.~J. Mele,
		  \href{http://dx.doi.org/10.1103/PhysRevLett.95.146802}{Phys. Rev. Lett.
		  \textbf {95} 146802,  (2005)}.
		
		\bibitem{PhysRevLett.95.226801}
		C.~L. Kane and E.~J. Mele,
		  \href{http://dx.doi.org/10.1103/PhysRevLett.95.226801}{Phys. Rev. Lett.
		  \textbf {95} 226801,  (2005)}.
		
		\bibitem{PhysRevLett.95.136602}
		L.~Sheng, D.~N. Sheng, C.~S. Ting, and F.~D.~M. Haldane,
		  \href{http://dx.doi.org/10.1103/PhysRevLett.95.136602}{Phys. Rev. Lett.
		  \textbf {95} 136602,  (2005)}.
		
		\bibitem{PhysRevLett.97.036808}
		D.~N. Sheng, Z.~Y. Weng, L.~Sheng, and F.~D.~M. Haldane,
		  \href{http://dx.doi.org/10.1103/PhysRevLett.97.036808}{Phys. Rev. Lett.
		  \textbf {97} 036808,  (2006)}.
		
		\bibitem{sheng_spin_2013}
		L.~Sheng, H.-C. Li, Y.-Y. Yang, D.-N. Sheng, and D.-Y. Xing,
		  \href{http://dx.doi.org/10.1088/1674-1056/22/6/067201}{Chin. Phys. B \textbf
		  {22} 067201,  (2013)}.
		
		\end{thebibliography}

\end{document}